\documentclass[pre,twocolumn]{revtex4-1}
\usepackage{psfrag}
\usepackage{amssymb,amsmath,amsthm}
\usepackage{graphicx}
\usepackage{verbatim}
\usepackage[colorlinks=true,linkcolor=blue,citecolor=blue]{hyperref}
\usepackage{subfigure}

\begin{document}

\title{Large deviation function and fluctuation theorem for classical particle
transport}
\author{Upendra Harbola}
\affiliation{Inorganic and Physical Chemistry, Indian Institute of Science, Bangalore}
\author{Christian Van den Broeck}
\affiliation{Hasselt University, B-3500 Hasselt, Belgium}
\author{Katja Lindenberg}
\affiliation{Department of Chemistry and Biochemistry and BioCircuits Institute,
University of California San Diego, La Jolla, CA 92093-0340}

\begin{abstract}
We analytically evaluate the large deviation function in a simple model of
classical particle transfer between two reservoirs. We illustrate how the
asymptotic large time regime is reached starting from a special propagating
initial condition.
We show that the steady state fluctuation theorem holds provided that the 
distribution of the particle number
decays faster than an exponential, implying analyticity of the generating function
and a discrete spectrum for its evolution operator. 
\end{abstract}

\maketitle
\section{Introduction}
The second law of thermodynamics has two major ingredients: the existence, in an
equilibrium state, of a state function called the entropy, 
and the increase of the total entropy in a spontaneous transition between two
equilibrium states.  The discovery of the fluctuation theorem entails a double
departure from this standard formulation of the second law. Entropy is defined
in nonequilibrium 
states, and it is even defined for a single realization  in a non-extensive
system, where it is a stochastic quantity which can increase as well as 
decrease with time. Furthermore, the fluctuation 
theorem  states (in its simplest formulation) that the fluctuating total entropy
production $\Delta S_{tot}$ obeys a symmetry property: 
$P(\Delta S_{tot})/P(-\Delta S_{tot})=\exp{\Delta S_{tot}/k_B}$ \cite{ft}. This
equality implies the second law inequality 
$\langle{\Delta S}_{tot}\rangle\geq 0$. 

The fluctuation theorem was first discovered in an ``asymptotic version,"
preceding the definition of a proper stochastic 
entropy \cite{udo}. In this version, one only focuses on the entropy production
in idealized reservoirs. For a heat reservoir $i$ 
at temperature $T^{(i)}
$, the entropy change is given by $\Delta S^{(i)}
=Q^{(i)}
/T^{(i)}
$,
where $Q^{(i)}
$ is the net amount of heat transferred from the system to reservoir $i$.
Note that $Q^{(i)}
$ and hence $\Delta S^{(i)}
=Q^{(i)}
/T^{(i)}
$ are well defined even though the system 
need not be at equilibrium. Furthermore, in the case of a small system, the
amount of heat $Q^{(i)}
$ will differ from one realization to 
another, so that this is a genuine stochastic quantity. 

Coming back to the
fluctuation theorem, one notes that the total entropy change 
is  the sum of all the contributions: $\Delta S_{tot}=\sum_i \Delta S^{(i)}
 +\Delta S$, 
which includes the entropy change $\Delta S$ of the system. 
Since the entropy productions in the reservoirs are easier to monitor than the
nonequilibrium stochastic system entropy $\Delta S$  (where, for simplicity, we
also avoid a discussion of the possible contribution to the entropy coming from
interaction terms between system and reservoirs), it is of interest  to identify 
situations in which the latter contribution is negligible. A good candidate  is
the asymptotic time regime for 
systems operating under steady state nonequilibrium conditions. Indeed one
intuitively expects that the heat (and/or particle) flows will be 
proportional to time, while the (stochastic) entropy of the system, being in a
steady state, should remain more or less constant. Hence the latter contribution
becomes negligible in the long-time limit. This was indeed proven to be the case
for a class of 
systems with bounded energy. One can then write 
$$
\frac{P(\sum_i\Delta S^{(i)}
)}{P(-\sum_i\Delta S^{(i)}
)}\sim\mbox{e}^{\sum_i\Delta S^{(i)}
/k_B},
$$
known as the steady-state fluctuation theorem \cite{RMP}.

This asymptotic  version of the fluctuation theorem can, however, break down when
the energy of the system is unbounded \cite{farago}. 
A well documented example is that of Brownian particle  in contact with one or
several heat reservoirs 
\cite{vanzon,visco,baiesi,puglisi,fogedby}. The large deviation function probes
exponentially unlikely events for the heat evacuated  
to the reservoirs. Such an event can however be the result of an exponentially
unlikely initial energy of the system. In this case the 
entropy contribution of the system is no longer negligible and a fluctuation
theorem in terms of reservoir entropies alone breaks down.

Here, we study the less-documented case of particle transport. More precisely, we
analyze particle transport through a system in 
contact with particle reservoirs \cite{vdb}. We evaluate analytically
the  large deviation function for the particle flux. 
We find that the asymptotic fluctuation theorem is satisfied under ``natural"
conditions, which are typical for the distribution of the number of particles.
More precisely, whereas the Boltzmann factor allows for large energies, albeit
in an exponentially unlikely way, the distribution $P(n)$ for having $n$
particles in a system typically decays faster than exponential, due to the
indistinguishability property giving rise to a $n!$ contribution in the
denominator. As a result the generating function $F(s)=\Sigma_n s^n P(n)$ is
analytic in $s$ in the entire complex plane, and the stochastic operator
describing the particle exchange has a discrete spectrum. Under this condition,
we are able to prove the validity of the steady state fluctuation theorem.

When the system is coupled to particle reservoirs, a cumulated particle flux develops 
between the system and the reservoirs. The probability distribution of this flux typically
 depends on time in a complicated way. One may wonder whether there exist special initial conditions
for which the functional form
of this distribution does not change with time. 
We will identify such an initial distribution for the model under consideration. 
Note that, from the point of view of a detector (observer), the natural initial condition is
to start to count the particle flux at time zero starting from a zero value.
However, as we shall see, this is an
unnatural initial condition from the ``system's point of view," as correlations then develop 
between the observed flux and the state of the system. 

The paper is organized as follows. In the next section we present the 
particle transfer model and its basic mathematical framework. In Sec.~\ref{GF}
we introduce the generating function method to compute  statistics 
of the 
particle fluxes between system and reservoirs. 
In Sec.~\ref{time} we identify
the joint distribution of system state and particle fluxes that 
propagates in time. In the long-time limit, it approaches an  asymptotic regime
characterized by a large deviation function. 
In Sec.~\ref{discussion} we verify the steady-state  fluctuation relation 
and discuss  its validity in the specific context of
particle transfer.

\section{Master equation}
We consider a system which can  exchange  matter (particles) with two particle
baths, with chemical potentials $\mu^{(1)}$ and $\mu^{(2)}$ and temperatures 
$T^{(1)}$ and $T^{(2)}$, respectively.  The number of particles in the system
will be denoted by  $n$, $ n\in \mathbb{N}
$. For simplicity, we will operate in the limit of non-interacting, classical
but indistinguishable particles, each of energy $\epsilon$. We furthermore assume
that the exchange of particles between system and baths can be described by a
Markovian jump process. Consider an instant at which the system contains $n$ particles.
Let $k_{+}^{(i)}$ be the rate for a particle to jump 
into the system, while $n k_{-}^{(i)}$ is the rate for a particle to jump out
of the system, to and from thermal bath $i$, with 
$i=1,2$. The corresponding master equation for
the probability to have $n$-particles in the system at time $t$ then reads
\begin{eqnarray}
 \label{eq-00}
\frac{\partial}{\partial t}P(n;t) &=& k_+ P(n-1;t)
 + (n+1)k_- P(n+1;t)  
 \nonumber\\
&-& (k_+ + nk_-)P(n;t),
\end{eqnarray} 
with
  \begin{equation}
  k_{\pm}=k_\pm^{(1)}+k_\pm^{(2)}.
  \end{equation}

We start with a number of remarks. First, when in contact with a single bath
$i$, the steady state probability solution of (\ref{eq-00})  reduces to the
Poissonian equilibrium distribution $P_{eq}^{(i)}(n)$,
 \begin{eqnarray}
 \label{eq-0}
P_{eq}^{(i)}(n) = \frac{\rho_i
^n}{n!} \mbox{e}^{-\rho_i},
\end{eqnarray}
with a corresponding average number of particles $\rho_i$ given by
\begin{equation}
\rho_i
=\frac{k_+^{(i)}}{k_-^{(i)}}.
\end{equation}
We mention for further use the relation between this equilibrium density and
the reservoir properties ($\beta^{(i)}=1/k_B T^{(i)}$),
\begin{equation}\label{0}
\rho_i=g e^{-\beta^{(i)}(\epsilon-\mu^{(i)})},
\end{equation}
where $g$ is the number of single particle states with energy $\epsilon$ in
the system and $\mu_i$ is the chemical potential of the $i$-th bath. 
Since we are dealing with the classical limit, $g$ should be
much larger than $n$ and the exponential in (\ref{0}) is much smaller
than unity. Equation (\ref{0}) is the detailed balance condition for the
transport between the system and the $i$-th reservoir.

Second, due to the combinatorial factor $n!$, the probability distribution 
$P_{eq}^{(i)}(n)$ decays more quickly than exponentially for large $n$. Since there is
a compelling physical reason for this factor, namely, the indistinguishability of
particles, we expect this to be a genuine feature of a particle distribution
function, and we will assume below a faster than exponential decay for the
probability distribution even when operating under nonequilibrium conditions. 

Third, we mention the following exact time-dependent solution of (\ref{eq-00}),
namely, a propagating Poisson distribution:
\begin{eqnarray}
 \label{eq-4}
P(n;t) = \frac{(\rho(t))^n \mbox{e}^{-\rho(t)}}{n!}
\end{eqnarray}
with the mean number of particles in the system at time $t$ given by
\begin{eqnarray}
 \label{eq-5}
\rho(t) = \frac{k_+}{k_-}+\left(\rho(0)-\frac{k_+}{k_-}\right)\mbox{e}^{-k_-t}.
\end{eqnarray}
At steady state,
$\rho_{ss}=k_+/k_-$. 
Hence,  the particle distribution maintains a Poissonian
equilibrium-like shape, even though it is in a nonequilibrium state.  
 
We next turn to our main quest, namely, the  study of the fluctuation theorem.
We first identify the entropy change $\Delta S_r^{(i)}$ in bath $(i)$, for a
given total elapsed time $t$:
\begin{eqnarray}
\label{ep}
\Delta S^{(i)}&=&\frac{ Q^{(i)}}{T^{(i)}}\nonumber\\
&=&-\frac{(\epsilon-\mu^{(i)}) N_i
}{T^{(i)}}\nonumber\\
&=&k_B N_i \ln \frac{\rho_i}{g}   
\end{eqnarray}
Here $N_i$ is a register that adds (subtracts) $1$ whenever a particles crosses from heat 
bath $i$ to the system (from the system to heat bath $i$). Thus, $N_i$ is the {\emph net} 
number crossing from bath $i$ to the system between time $0$ and $t$ {\emph plus} the 
number that was on the register at time $t=0$ (see below).
All the particles have the same energy $\epsilon$. 
In going from the first to the second line, we have
used the conservation of energy, with the change in 
bath energy $-\epsilon N_i$ being
equal to heat plus chemical energy, 
$Q^{(i)} -\mu^{(i)} N_i$. Transition to the
third line is based on (\ref{0}).

The evaluation of the stochastic bath entropies is thus reduced to that of the
number of particles $N_{i}$ ($  N_i \in  \mathbb{Z}, i=1,2$)
transferred from baths to
the system in time $t$. Since these numbers are deterministic
functions of the system dynamics, the enlarged set of variables $n,N_1,N_2$
again defines a Markov jump  process, and 
the joint probability $P(n,N_1,N_2;t)$, to find $n$ particles in the system
while 
having a cumulative transfer of $N_1,N_2$ particles in time $t$, evolves according to the
following master equation:
\begin{eqnarray}
 \label{eq-2}
\frac{\partial}{\partial t}P(n,N_1,N_2;t) &=& k_+^{(1)}
P(n-1,N_1-1,N_2;t)\nonumber\\
 &+&  k_+^{(2)} P(n-1,N_1,N_2-1;t) \nonumber\\
 &+& (n+1)k_-^{(1)} P(n+1,N_1+1,N_2;t)  \nonumber\\
&+& (n+1) k_-^{(2)} P(n+1,N_1,N_2+1;t) \nonumber\\
&-& (k_+ + nk_-)P(n,N_1,N_2;t).
\end{eqnarray}

As we proceed to show, it is possible to find an exact time-propagating
solution of this equation, which allows us to find the asymptotic large time
properties, in particular those of the stochastic entropy. This solution furthermore
illustrates how this asymptotic regime is reached in the course of time.

We finally note that due to particle conservation, the following identity  holds 
at all times:
\begin{eqnarray}
 \label{par-cons}
n(t)= N_1(t)-N_1(0)+N_2(t)-N_2(0) + n(0),
\end{eqnarray}
where $n(0)$, $N_1(0)$ and $N_2(0)$ are, respectively, the number of particles 
in the system and the number of particles on registers $1$ and $2$
at time $t=0$. We make two observations whose significance will  reveal themselves
when discussing the time-propagating solution of (\ref{eq-2}). 
First, while it is quite natural and tempting, from the observer's point of view, to choose $N_1(0)=N_2(0)=0$,
the choice of $N_1(0)$ and $N_2(0)$ is in 
principle free, and could even be stochastic. 
Second, it follows from (\ref{par-cons}) that the condition  $n(0)=N_1(0)+N_2(0)$ propagates in time, i.e., it implies 
that $n(t)=N_1(t)+N_2(t)$ holds at all times.

\section{Generating function}
\label{GF}
The solution of the master equation
is facilitated by switching to the following generating 
function:
\begin{eqnarray}
 \label{eq-6}
F_{\boldsymbol \lambda}(s;t) = \sum_{N_1,N_2=-\infty}^\infty \;\sum_{n=0}^\infty \;
\mbox{e}^{\lambda_1 N_1+\lambda_2N_2}\;s^n\; 
P(n,N_1,N_2;t).\nonumber\\
\end{eqnarray}
The parameters  ${\boldsymbol \lambda}=\{\lambda_1,\lambda_2\}$ are so-called ``counting parameters"
that keepi track of the net number of particles transferred between the system and
the corresponding reservoirs. We shall use $\boldsymbol \lambda$ to denote 
a dependence on $\lambda_1$ and $\lambda_2$.

By combination with (\ref{eq-2}) we find
\begin{eqnarray}
 \label{eq-7}
\frac{\partial}{\partial t} F_{\boldsymbol \lambda}(s;t) &=& {\cal L}_{\boldsymbol
\lambda}(s)F_{\boldsymbol \lambda}(s,0),
\end{eqnarray}
 where the operator ${\cal L}$ is defined as,
\begin{eqnarray}
 \label{eq-8}
{\cal L}_{\boldsymbol \lambda}(s) = k_- \alpha_{\boldsymbol \lambda} s 
+ k_- \beta_{\boldsymbol \lambda}
\frac{\partial}{\partial s} - k_- s\frac{\partial}{\partial s} -k_+
\end{eqnarray}
with 
\begin{eqnarray}
\label{a-b}
\alpha_{\boldsymbol \lambda} &=&
\frac{k_+^{(1)}\mbox{e}^{\lambda_1}+k_+^{(2)}\mbox{e}^{\lambda_2}}{k_-},\\  
\beta_{\boldsymbol \lambda} &=&
\frac{k_-^{(1)}\mbox{e}^{-\lambda_1}+k_-^{(2)}\mbox{e}^{-\lambda_2}}{k_-}. 
\end{eqnarray}
Since the dependence  on the counting parameters in  Eq.~(\ref{eq-7}) is
parametric, it suffices to evaluate  the eigenvectors and eigenvalues of  ${\cal
L}$ as an operator with respect to the variable $s$.
Let $\Psi_{\boldsymbol \lambda}(s)$ be an eigenvector of 
${\cal L}_{\boldsymbol \lambda}(s)$
with eigenvalue $\zeta_{\boldsymbol \lambda}$,
\begin{eqnarray}
\label{eq-9}
{\cal L}_{\boldsymbol \lambda}(s) \Psi_{\boldsymbol \lambda}(s) = \zeta_{\boldsymbol \lambda} 
\Psi_{\boldsymbol \lambda}(s).
\end{eqnarray}
With the expression (\ref{eq-8}) for the operator, the eigenfunctions are found
by straightforward integration,
\begin{eqnarray}
 \label{eq-10}
\Psi_{\boldsymbol \lambda}(s) = \left(s-\beta_{\boldsymbol \lambda}\right)^{g_{\boldsymbol \lambda}}
\mbox{exp}
\left\{\alpha_{\boldsymbol \lambda}\left(s-\beta_{\boldsymbol \lambda}\right)\right\},
\end{eqnarray}
where
\begin{eqnarray}
 \label{eq-10a}
g_{\boldsymbol \lambda} = \alpha_{\boldsymbol \lambda}\beta_{\boldsymbol \lambda}-\frac{k_+ 
+ \zeta_{\boldsymbol \lambda}}{k_-}.
\end{eqnarray}

We now make the following crucial assumption, already mentioned in the
introduction: 
We request that the  eigenfunctions be analytic in the variable $s$ in the entire
complex plane.  Analyticity imposes two restrictions on the exponent 
$g_{\boldsymbol \lambda}$:
it must be greater than or equal to zero, $g_{\boldsymbol \lambda} \geq 0$ and it must
be an integer. Hence, setting $g_{\boldsymbol \lambda} =l$ with $l\in \mathbb{N}$
we find from (\ref{eq-10a}) 
for the eigenspectrum of the operator ${\cal L}_{\boldsymbol \lambda}$
\begin{eqnarray}
 \label{eq-11}
\zeta^{(l)} _{\boldsymbol \lambda} = k_-\alpha_{\boldsymbol \lambda}\beta_{\boldsymbol \lambda}-k_+ -
lk_-,\;\;\; l\in \mathbb{N}.
\end{eqnarray} 
The corresponding eigenfunction can now be written in the following compact way:
\begin{eqnarray}
 \label{eq-11a}
\Psi_{\boldsymbol \lambda}^{(l)}(s) = \frac{\partial^l}
{\partial \alpha^l_{\boldsymbol \lambda}} 
\mbox{exp} \left\{\alpha_{\boldsymbol \lambda}\left(s-\beta_{\boldsymbol
\lambda}\right)\right\}.
\end{eqnarray}
At this point, we make two observations. First, the eigenvalue
$\zeta^{(l)}_{\boldsymbol \lambda}
$ depends on ${\boldsymbol \lambda}$ only via  $\lambda=\lambda_1-\lambda_2$ and is
invariant under the following interchanges:
\begin{eqnarray}
\label{eq-12}
\lambda \leftrightarrow{-\lambda-\ln
\frac{\rho_1}{\rho_2}}\\
\lambda_i\leftrightarrow-\lambda_i-\ln
(\rho_i)\;\;\;\mbox{for}\;i=1 \;\mbox{and}\; 2.
\end{eqnarray}
The eigenfunctions themselves, however, do not obey this symmetry. This property
will  be crucial to verify the steady state fluctuation theorem.

\begin{figure}[h]
\centering
\mbox{\subfigure{\includegraphics[width=1.7in,height=1.4in]{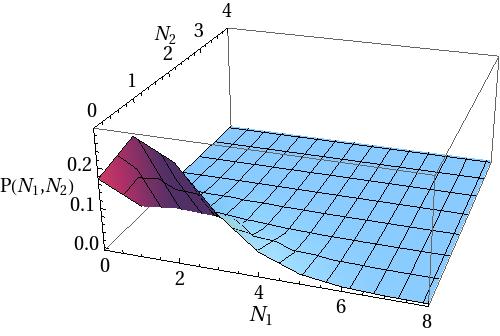}}
\hspace{.1cm}
\subfigure{\includegraphics[width=1.65in,height=1.4in]{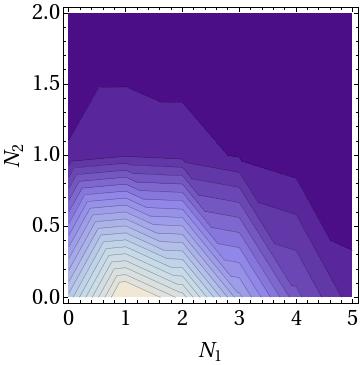} }}
\mbox{\subfigure{\includegraphics[width=1.7in,height=1.4in]{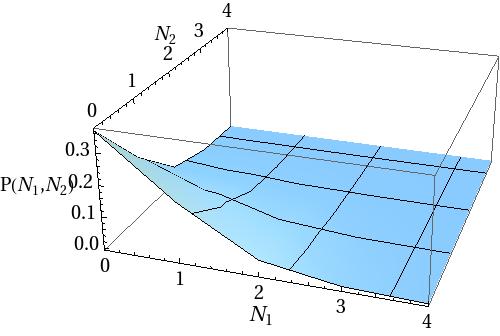}}
\hspace{.1cm}
\subfigure{\includegraphics[width=1.65in,height=1.4in]{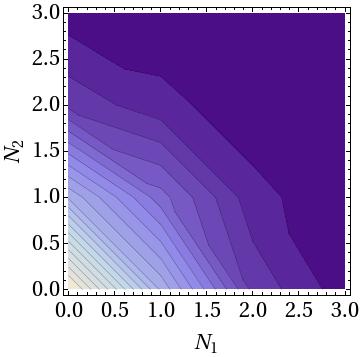} }}
\caption{(Color online) Upper panel: Initial distribution $P(N_1,N_2;t=0)$, cf. Eq. (\ref{eq-17x1}), and its projection
onto the $(N_1,N_2)$ plane. Parameters
are: $k_+^{(1)}=2.0, k_-^{(1)}=0.2, k_+^{(2)}=0.1$ and $k_-^{(2)}=1.0$.
Lower panel: Initial probability distribution correspondng to  equilibrium values
$k_+^{(1)}= k_-^{(1)}= k_+^{(2)}=k_-^{(2)}=1$.
}
 \label{fig-1}
\end{figure}

Second, the above set of eigenfunctions is complete in the sense that any
analytic function of $s$  can be expanded in terms of this basis. Furthermore,
the expansion is unique as it corresponds to a Taylor expansion around the point
$\beta_{\bf \lambda}$. Hence  we obtain the following explicit expression for
the generating function (assumed to be analytic in $s$) obeying (\ref{eq-7}):
\begin{eqnarray}
\label{eq-17}
F_{\boldsymbol \lambda}(s;t) = \sum_{l=0}^\infty a^{(l)}_{\boldsymbol\lambda}\;
\Psi_{\boldsymbol \lambda}^{(l)}(s)\; \mbox{e}^{t\zeta^{(l)}_{\boldsymbol \lambda}},
\end{eqnarray}
where $a^{(l)}_{\boldsymbol\lambda}$ is expansion coefficient of the eigenfunction 
$\Psi_{\boldsymbol \lambda}^{(l)}(s)$ for the initial function 
$F_{\boldsymbol \lambda}(s,0)$. 
In the sequel, we will focus on a particular simple initial condition,
corresponding to $a^{(l)}_{\boldsymbol\lambda}=\delta^K_{l,0} a^{(0)}_{\boldsymbol\lambda}$, 
where $\delta^{K}$ is the
Kronecker delta. One reason is obvious: the
corresponding eigenfunction is dominating the long-time limit, since it has the
lowest eigenvalue, cf. (\ref{eq-11}). The other reason is that such an initial condition corresponds, for  an appropriate choice
of the coefficient $a^{(0)}_{\boldsymbol\lambda}$, to a genuine probability distribution. 
The explicit expression for $\Psi_{\boldsymbol \lambda}^{(0)}(s)$ 
suggests the following choice for $a^{(0)}_{\boldsymbol\lambda}$: 
\begin{equation}\label{a0}
a^{(0)}_{\boldsymbol\lambda}=\mbox{e}^{{\alpha}_{\boldsymbol\lambda}\beta_{\boldsymbol\lambda}-k_+/k_-},
\end{equation}
where the subtraction of $k_+/k_-$ guarantees normalization.
Referring to the appendix for the calculation of the inverse,  
we find that it leads  to the following initial
probability distribution:
\begin{eqnarray}
\label{distribution}
P(n,N_1,N_2;t=0)&=&
\frac{\mbox{e}^{-\frac{k_+}{k_-}}}{N_1!N_2!}
\left(\frac{k_+^{(1)}}{k_-}\right)^{N_1}
\left(\frac{k_+^{(2)}}{k_-}\right)^{N_2}\nonumber\\
&\times&\delta^K_{n,N_1+N_2}\Theta(N_1)\Theta(N_2),
\end{eqnarray}
where $\Theta(x)$ is a Heaviside theta-function.
The Kronecker delta in (\ref{distribution})
imposes the condition that initially $n(0)=N_1(0)+N_2(0)$. 
This condition was to be expected since, as mentioned earlier, it 
propagates in time with $n=N_1+N_2$ at all times. 
The corresponding reduced distribution for the number of particles in the
system, $P^{st}(n)$, is, as expected, the steady state distribution, cf. (\ref{eq-4}),
 \begin{eqnarray}
 \label{eq-4eq}
P^{st}(n) \equiv  P(n;t=0) = \frac{\rho^n \mbox{e}^{-\rho}}{n!},
\end{eqnarray}
with $\rho=k_+/k_-$. 
 The reduced distribution for the initial
cumulated particle transfer  $P(N_1,N_2;t=0)$, see also Fig.~\ref{fig-1},
 is obtained by summation of  (\ref{distribution}) over $n$. The summation 
only affects the Kronecker delta; hence $P(N_1,N_2;t=0)$ is obtained from
 Eq. (\ref{distribution}) by the replacement of the Kronecker delta
$\delta^K_{n,N_1+N_2}$ with $\Theta(N_1+N_2)$, but this factor is superfluous due to the presence of
$\Theta(N_1)\Theta(N_2)$. We conclude that the reduced initial distribution $P(N_1,N_2;t=0)$ is a product
of two independent Poissonian distributions. One can verify that the ``natural initial condition" $P(N_1,N_2;t=0)= \delta^{K}_{N_1,0} \delta^{K}_{N_2,0}$ does not lead to a time propagating solution, that is, it cannot be expressed solely in terms of the eigenvector $\Psi_{\boldsymbol \lambda}^{(0)}(s)$.
This could have been anticipated from the fact that the propagating condition $n(0)=N_1(0)+N_2(0)$ would then imply $n(0)=0$, which is incompatible with the 
"propagating" steady state statistics $P^{st}(n)$ for $n$.

\section{Propagating solution and large deviation function}
\label{time}

\begin{figure}[h]
\centering
\mbox{\subfigure{\includegraphics[width=1.7in,height=1.4in]{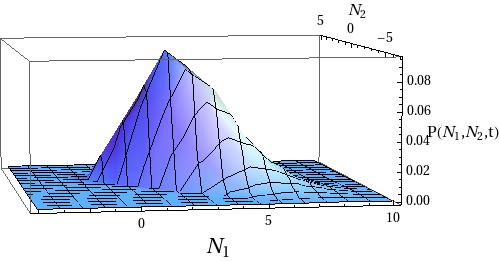}}
\hspace{.1cm}
\subfigure{\includegraphics[width=1.65in,height=1.4in]{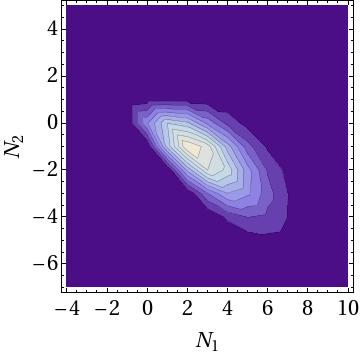} }}
\mbox{\subfigure{\includegraphics[width=1.7in,height=1.4in]{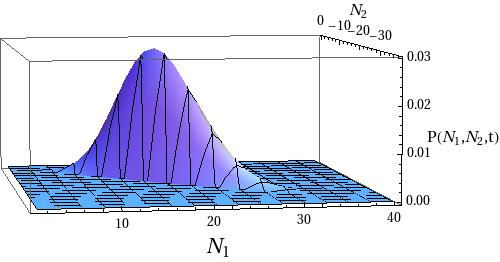}}
\hspace{.1cm}
\subfigure{\includegraphics[width=1.7in,height=1.4in]{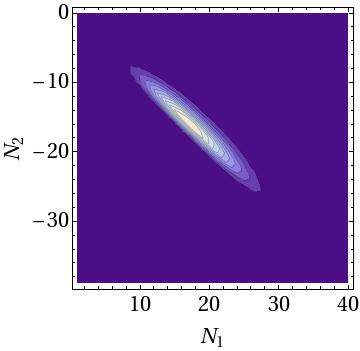} }}
\mbox{\subfigure{\includegraphics[width=1.7in,height=1.4in]{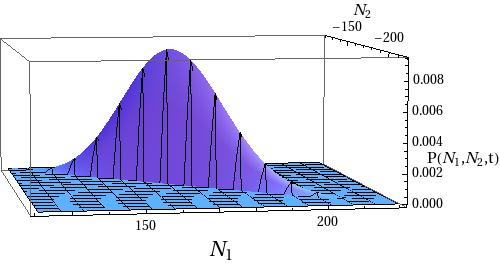}}
\hspace{.1cm}
\subfigure{\includegraphics[width=1.7in,height=1.4in]{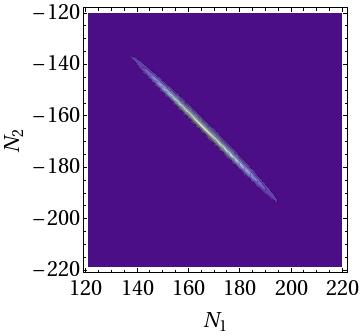} }}
\caption{(Color online) Probability distribution function, $P(N_1,N_2;t)$, for 
different times, $t=1,10$ and $100$ (top to bottom).  As  time  increases, the 
distribution becomes more peaked around the second diagonal, 
illustrating the convergence of $j_1=N_1/t$ 
to $-j_2=-N_2/t$.}
 \label{fig-1a}
\end{figure}

Our main focus is the evaluation of the joint reduced distribution, 
$P(N_1,N_2;t)$. 
Its  generating function $F_{\boldsymbol \lambda}(t)$  is found 
by setting $s=1$ in (\ref{eq-17}):  
\begin{eqnarray}
 \label{eq-17a}
F_{\boldsymbol \lambda}(t) &=& \sum_{N_1,N_2=-\infty}^\infty  \mbox{e}^{\lambda_1
N_1+\lambda_2N_2} 
P(N_1,N_2;t)\\
&=& \sum_{l=0}^{\infty}  a^{(l)}_{\boldsymbol\lambda}\;
\Psi_{\boldsymbol \lambda}^{(l)}(s=1)\;\mbox{e}^{t\zeta^{(l)}_{\boldsymbol
\lambda}}.\label{eq-17b}
\end{eqnarray}
The  $l=0$ term dominates  the series in Eq.~(\ref{eq-17}) for asymptotically long times 
$t \rightarrow \infty$. Alternatively, this term corresponds to the full 
solution at all times for the initial condition identified in the previous
section. We henceforth consider this case and can thus write
\begin{eqnarray}
 \label{eq-17b}
F_{\boldsymbol \lambda}(t)= a^{(0)}_{\boldsymbol\lambda} \mbox{e}^{t\zeta^{(0)}_{{\boldsymbol \lambda}}} 
\Psi^{(0)}_{\boldsymbol\lambda}(s=1),
\end{eqnarray}
with $a^{(0)}_{\boldsymbol\lambda} $ given by Eq. (\ref{a0}).
The corresponding joint probability $P(N_1,N_2;t)$ is computed by taking the
inverse transform of (\ref{eq-17b}): 
\begin{eqnarray}
 \label{eq-17c}
P(N_1,N_2;t) &=& \oint \frac{d\lambda_1}{2\pi i} \oint \frac{d\lambda_2}{2\pi i} 
\mbox{e}^{-(\lambda_1 N_1+\lambda_2 N_2-t\zeta^{(0)}_{{\boldsymbol \lambda}})} \nonumber\\
&\times& a^{(0)}_{\boldsymbol\lambda} \Psi_{\boldsymbol\lambda}^{(0)}(s=1).
\end{eqnarray} 
Switching  to the integration variable $\lambda=\lambda_1-\lambda_2$, and
expanding the exponential of $\alpha_{\boldsymbol\lambda}/k_-$ which appears inside $\Psi_{\boldsymbol \lambda}^{(0)}(s=1)$, cf. 
Eq.~(\ref{eq-11a}), Eq. (\ref{eq-17c})  can be
rewritten as
\begin{eqnarray}
 \label{eq-17x1}
&&P(N_1,N_2;t) = \sum_{m=0}^\infty \frac{1}{m!}\frac{1}{k_-^m}
\oint \frac{d\lambda_1}{2\pi i} \mbox{e}^{\lambda_1(m-N_1-N_2))}\nonumber\\
&&\times \oint \frac{d\lambda}{2\pi i} (k_+^{(1)}+k_+^{(2)}\mbox{e}^{-\lambda})^m 
\mbox{e}^{\lambda
(N_2+t\zeta^{(0)}_\lambda)}\mbox{e}^{-k_+/k_-} .
\end{eqnarray}
Since $\zeta^{(0)}_{\lambda}$ and $\alpha_\lambda \beta_\lambda$
are functions of only $\lambda$, the
integral over $\lambda_1$ reduces to the Kronecker delta 
$\delta^K_{m,N_1+N_2}$. 
Expanding the  exponential in $\alpha_\lambda \beta_\lambda$, the remaining
integral over $\lambda$ can be performed. Following the same steps which 
led to (\ref{distribution}), we obtain the following propagating solution for the 
joint distribution function:
\begin{eqnarray}
 \label{eq-17x2}
P(N_1,N_2;t) &=&
\frac{\mbox{e}^{-t \frac{k_+^{(1)}k_-^{(2)}+k_+^{(2)}k_-^{(1)}}{k_-}}}
{(N_1+N_2)!} \mbox{e}^{-k_+/k_-}\nonumber\\
&\times& \left(\frac{x}{2k_-^{(2)}}\right)^{N_2}
\left(\frac{k_+^{(1)}}{k_-}\right)^{N_1} \nonumber\\
&\times&\sum_{m=0}^{N_1+N_2} \binom{N_1+N_2}{m} 
\left(\frac{k_+^{(2)}k_-^{(2)}}{k_+^{(1)}k_-^{(1)}}\right)^{m/2} \nonumber\\
&\times& I_{m-N_2}\left(xt\right)\Theta(N_1+N_2),
\end{eqnarray}
where 
\begin{eqnarray}
 \label{x-eq}
x=\frac{2}{k_-}\sqrt{k_+^{(1)}k_+^{(2)}k_-^{(1)}k_-^{(2)}}
\end{eqnarray}
and $I_n(y)$ is the modified bessel function of the first kind 
of order $n$. It is straightforward to check that for $t=0$ this 
reduces to the product of Poissonians, cf. (\ref{distribution}).
The distribution function (\ref{eq-17x2}) is shown in Fig.~\ref{fig-1a} for
different values of $t$. 


We next focus on  the large $t$ limit.  Since the particle fluxes
$N_1$ and $N_2$ diverge for $t \rightarrow \infty$, we introduce the fluxes per unit time 
$j_i=N_i/t$.
In this limit, an additional simplification takes place as one finds asymptotically
that the stochastic quantities $j_1$ and $j_2$ become identical (see Fig.~\ref{fig-1a}).
Hence, the statistics in the long time limit are
expressed in terms of a single flux $j_1=-j_2=j$.
 The corresponding probability distribution function for $j$ 
has the typical shape from large deviation theory,  namely,

\begin{equation}
\label{pdf-t}
P(j;t) \sim \mbox{e}^{-t{\cal L}(j)}
\end{equation}
with the
large deviation function
\begin{eqnarray}
 \label{eq-17f}
{\cal L}(j) = -\lim_{t\to \infty}\frac{1}{t}\ln
P(j;t)
\end{eqnarray}
given by the following convex non-negative function:
\begin{eqnarray}
 \label{eq-17g}
{\cal L}(j) &=&  \frac{k_+^{(1)}k_-^{(2)}+k_+^{(2)}k_-^{(1)}}{k_-}
-\sqrt{x^2+j^2}\nonumber\\
&+&\frac{j}{2}\ln
\left(\frac{\rho_2}{\rho_1}\frac{\sqrt{x^2+j^2}+j}{\sqrt{x^2+j^2}-j}\right).
\end{eqnarray}

\begin{figure}[h]
\centering
\rotatebox{0}{\includegraphics[width=2.5in]{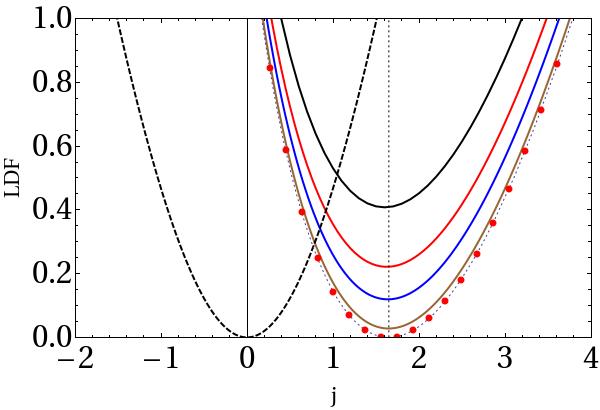}}
\caption{(Color online) The function $-\ln P(N;t)/t$, and its limiting form,  the large deviation function (\ref{eq-17f}),
${\cal L}(j)$,  as a function of time $t=10,20,40,200$ (solid curves,top to
bottom). $j=N/t$ is plotted along the x-axis.
The parameter values are the same as in
Fig.~\ref{fig-1}. The dots represent the analytical result in Eq. (\ref{eq-17g})
and the dotted curve is a guide to the eye.
The dashed curve with a minimum at $j=0$ represents the equilibrium case ($\rho_1=\rho_2$)
for $k_+^{(1)}=k_-^{(1)}=k_+^{(2)}=k_-^{(2)}=1$. The dotted vertical line 
represents the average $\langle j\rangle$, Eq. (\ref{eq-19}).
}
 \label{fig-2}
\end{figure}

From Eq. (\ref{eq-17x2}), the probability $P(N,t)$ to have $N_1=-N_2=N$ at time $t$
takes a simple form,
\begin{eqnarray}
 \label{pnt}
P(N;t) &=&
\mbox{e}^{-t \frac{k_+^{(1)}k_-^{(2)}+k_+^{(2)}k_-^{(1)}}{k_-}}
\left(\frac{k_+^{(1)}k_-^{(2)}}{k_+^{(2)}k_-^{(1)}}\right)^{N/2}\nonumber\\
 &\times&\mbox{e}^{-k_+/k_-} I_{N}\left(xt\right).
\end{eqnarray}
In Fig.~\ref{fig-2} we plot $-(1/t)\ln P(N,t)$ to illustrate how 
the asymptotic form of the large deviation function is approached in the
course of time. The dots represent the exact analytical result, Eq. (\ref{eq-17g}).

An alternative procedure to obtain the large deviation function is to 
work via the scaled cumulant generating function,
\begin{equation}
\label{cumulant-GF}
G_{\boldsymbol\lambda}=\frac{1}{t}\ln F_{\boldsymbol\lambda}(t).
\end{equation}
For large $t$, the generating function $F_{\boldsymbol\lambda}(t)$ in Eq. (\ref{eq-17b}) can 
be approximated by
\begin{eqnarray}
 \label{approx-gf}
F_{\lambda}(t) = \mbox{e}^{t\frac{k_+^{(1)}k_-^{(2)}}{k_-}(\mbox{e}^\lambda-1)
+t\frac{k_+^{(2)}k_-^{(1)}}{k_-}(\mbox{e}^{-\lambda}-1)},
\end{eqnarray}
which on substituting in (\ref{cumulant-GF}) gives
\begin{eqnarray}
 \label{approx-cgf}
G_{\lambda} = \frac{k_+^{(1)}k_-^{(2)}}{k_-}(\mbox{e}^\lambda-1)
+\frac{k_+^{(2)}k_-^{(1)}}{k_-}(\mbox{e}^{-\lambda}-1).
\end{eqnarray}
$G_\lambda$ is related to ${\cal L}(j)$ by a Legendre transformation,
\begin{eqnarray} 
\label{LDF-new}
{\cal L}(j) = ext_{\lambda}\left(\lambda j - G_{\lambda}\right).
\end{eqnarray}
The extremum is found at
\begin{eqnarray}
 \label{eq-18}
\lambda(j) = \frac{1}{2}\ln\left(\frac{\rho_2}{\rho_1}
\frac{\sqrt{j^2+x^2}+j}{\sqrt{j^2+x^2}-j}\right).
\end{eqnarray}
Substituting this result in (\ref{LDF-new}), we recover Eq. (\ref{eq-17g}).

The cumulant generating function $G_\lambda$ allows for a swift calculation 
of the cumulants of the current. For large times, the $n$th cumulant 
$\kappa_n(N_1)$ of $N_1$, the net number of particles
transferred between the system and the reservoir by time $t$, 
is obtained from $G_\lambda$ as
\begin{eqnarray}
\label{cumulant-def}
\kappa_n(N_1) = t \left.\frac{d^n G_{\lambda}}{d\lambda^n}\right|_{\lambda=0}.
\end{eqnarray}
Thus according to (\ref{cumulant-def}), all cumulants  $\kappa_n (N_1)$ vary linearly 
with time. For particle current $j$, the average and the variance are given by:
\begin{eqnarray}
 \label{eq-19}
\kappa_1(j) &=&
\frac{1}{k_-}\left(k_+^{(1)}k_-^{(2)}-k_+^{(2)}k_-^{(1)}\right),\\
\kappa_2(j) &=&
\frac{1}{tk_-}\left(k_+^{(1)}k_-^{(2)}+k_+^{(2)}k_-^{(1)}\right).
\end{eqnarray}
\vspace{.4cm}

\begin{figure}[h]
\centering
\rotatebox{0}{\includegraphics[width=2.5in]{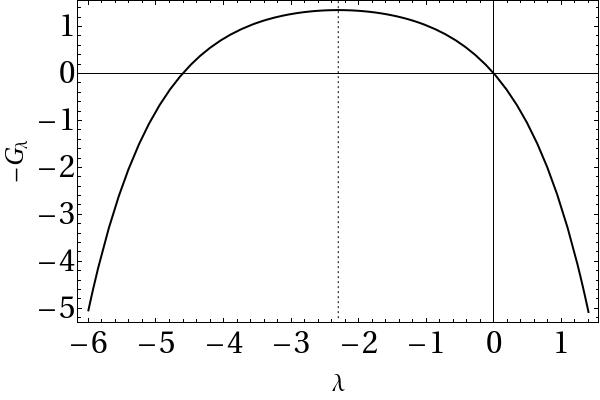}}
\caption{The cumulant generating function $G_\lambda$ for the same parameter values as in Fig.~\ref{fig-1}. 
Note the symmetry around $\lambda=(1/2)\ln(\rho_2/\rho_1)$ (dotted line), in agreement with 
the fluctuation theorem, cf. Eq. (\ref{eq-24})}.
 \label{fig-3}
\end{figure}

Furthermore, all odd and even cumulants of $j$
 are proportional to the first and second cumulants, respectively: 
\begin{eqnarray}
 \label{eq-21}
\kappa_n(j) &=& \frac{\kappa_2(j)}{t^{n-2}},~~~ n~~~ \mbox{even}\nonumber\\
\kappa_n(j) &=& \frac{\kappa_1(j)}{t^{n-1}},~~~ n~~~ \mbox{odd}.
\end{eqnarray}

\section{Fluctuation Theorem}
\label{discussion}

From (\ref{ep}), we find that the total entropy production in the reservoirs
(divided by $k_B$) is asymptotically given by
\begin{eqnarray}\label{eq-22}
\frac{\Sigma_i \Delta S^{(i)}}{k_B}
&=& \Sigma_i N_i  \ln \frac{\rho_i}{g} \\
&\sim &  t j \ln\frac{\rho_1}{\rho_2},
\end{eqnarray}
where we used the fact that asymptotically $j=j_1=-j_2$ ($N_i=t j_i$).
The steady state fluctuation theorem then requires that
\begin{equation}
\frac{P(j)}{P(-j)}\sim \mbox{e}^{t j \ln\frac{\rho_1}{\rho_2}},
\end{equation}
or, more precisely, that the large deviation function, Eq.~(\ref{eq-17g}), obeys
the symmetry relation,
\begin{equation}
\label{eq-23}
{\cal L}(j)-{\cal L}(-j)=  j \ln(\rho_2/\rho_1),
\end{equation}
which is 
easily verified by Eq. (\ref{eq-17g}). This symmetry of the large deviation function 
implies an analogous symmetry for the
cumulant generating function, 
\begin{equation}
\label{eq-24}
G_\lambda =G_{-\lambda-\ln(\rho_1/\rho_2)}.
\end{equation}

We recall that the  analyticity of the generating function for the particle number $n$ is 
an essential assumption in the above derivation of  the steady state fluctuation theorem. 
This requires that the corresponding probability $P(n)$ decay 
faster than an exponential in $n$. 
This property is verified by the steady state
Poisson distribution, Eq.~(\ref{eq-4}), which
decays (for large $n$) logarithmically faster than the exponential: 
$
P(n) \sim \mbox{e}^{-n\ln(n)},~~ \mbox{for}~ n\gg1 
$.
It is therefore quite natural to assume that the initial condition satisfies the same property, 
that is, that it ask decays faster than an exponential. 
Without this assumption, exponentially rare fluctuations in the initial particle distribution 
will lead to a breakdown of the fluctuation theorem.  


We finally mention that the large deviation function Eq.~(\ref{eq-17g})  is identical to that
 for a random walker on a line  with jump rates
${k_+^{(1)}k_-^{(2)}}/{k_-}$ to the right and
${k_+^{(2)}k_-^{(1)}}/{k_-}$ to the left. The physical interpretation is clear. Since the probability distribution of the
number of particles
contained in the system decays faster than an exponential, the large deviation statistics is essentially described
by the transfer statistics between the reservoirs only, cf. the asymptotic identity of $j_1$ and $-j_2$. It is intuitively clear that this long-time process will be identical  to the asymptotic properties of a random walk. 
 An  identical result is obtained for the effusion of particles between two reservoirs connected through a small opening \cite{effusion}. 
 
\section*{Acknowledgments}
UH acknowledges  financial support from the Indian Institute of Science, Bangalore, India. CVdB thanks the European Science Foundation through the network ``Exploring the Physics of Small Devices."  KL gratefully acknowledges support of the Office of Naval Research through Grant No. N00014-13-1-0205.
.....

 \section*{Appendix}
\label{appendix}

\section*{ Derivation of Eq. (\ref{distribution})}

In order to compute the initial probability distribution for
which only the term with $l=0$ in the series (\ref{eq-17}) 
survives, we need to inverse transform 
$a^{(0)}_{\boldsymbol\lambda}\Psi_\lambda^{(0)}(s)$, where 
we choose $a^{(0)}_{\boldsymbol\lambda}=
\mbox{e}^{{\alpha}_{\boldsymbol\lambda}\beta_{\boldsymbol\lambda}-k_+/k_-}$.
Note that for $\boldsymbol\lambda=0$, $a^{(0)}_{\boldsymbol\lambda}=1$
to preserve the normalization of the probability distribution. This is 
not the only possible choice for $a^{(0)}_{\boldsymbol\lambda}$, however this 
choice leads to a simple natural initial Poissoinian distribution, cf. Eq. (\ref{distribution}),
which propagates in time.  
Thus we have,
\begin{eqnarray}
 \label{app-1}
P(n,N_1,N_2;t=0) &=& \oint \frac{ds}{2\pi i} s^{n-1} 
\oint \frac{d\lambda_1}{2\pi i} \mbox{e}^{-\lambda_1N_1} \nonumber\\
&& \oint \frac{d\lambda_2}{2\pi i}  \mbox{e}^{-\lambda_2N_2} 
a^{(0)}_{\boldsymbol\lambda} \Psi_{\boldsymbol \lambda}^{(0)}(s).
\end{eqnarray}
Substituting for $\Psi_{\boldsymbol\lambda}^{(0)}(s)$ from Eq. (\ref{eq-11a}), we get
\begin{eqnarray}
 \label{app-2}
P(n,N_1,N_2;t=0) &=& \oint \frac{ds}{2\pi i}\frac{1}{s^{n+1}} 
\oint \frac{d\lambda_1}{2\pi i} \mbox{e}^{-\lambda_1N_1} \nonumber\\
&&\oint \frac{d\lambda_2}{2\pi i} 
\mbox{e}^{-\lambda_2N_2}
\mbox{e}^{s\alpha_{\boldsymbol\lambda}}
\mbox{e}^{-k_+/k_-}
\end{eqnarray}
Next we expand the exponential which contains the variable $s$. This allows us to
perform the $s$-integral and gives the Kronecker delta function $\delta^K_{n,m}$. 
We find
\begin{eqnarray}
 \label{app-3}
P(n,N_1,N_2;t=0) &=& \frac{1}{n!} \oint \frac{d\lambda_1}{2\pi i} \mbox{e}^{-\lambda_1N_1}
\oint \frac{d\lambda_1}{2\pi i} \alpha_\lambda^n\nonumber\\
&& \mbox{e}^{-\lambda_2N_2} 
\mbox{e}^{-k_+/k_-}.
\end{eqnarray}
Using $\alpha_{\boldsymbol\lambda}$ from (\ref{a-b}), and
expanding $\alpha_{\boldsymbol\lambda}^n$ using binomial expansion, 
we obtain
\begin{eqnarray}
 \label{app-3}
P(n,N_1,N_2;t=0) &=& \frac{\mbox{e}^{-\frac{k_+}{k_-}}}{n!}
\left(\frac{k_+^{(1)}}{k_-}\right)^n
\sum_{l=0}^n \binom{n}{l} \left(\frac{k_+^{(2)}}{k_+^{(1)}}\right)^l
\nonumber\\
&\times& \oint \frac{d\lambda_1}{2\pi i} \mbox{e}^{\lambda_1(n-l-N_1)}
\oint \frac{d\lambda_1}{2\pi i} \mbox{e}^{\lambda_2(l-N_2)}.\nonumber\\
\end{eqnarray}
The integrals over $\lambda_1$ and $\lambda_2$ give Kronecker deltas,
$\delta^{K}_{n,N_1+N_2}$ and $\delta^{K}_{l,N_2}$, respectively. 
Using these in (\ref{app-3}), we get (for $n=N_1+N_2$) Eq. (\ref{distribution}).

An alternative method to obtain the probability distribution is to
expand the generating function $F_{\boldsymbol\lambda}$ and compare
it term-by-term with the definition (\ref{eq-6}). Here we present this method
to recover Eq. (\ref{distribution}). At $t=0$, the generating function 
$F_{\boldsymbol\lambda}$ is (keeping only the $l=0$ term)
\begin{eqnarray}
\label{gf-n1}
 F_{\boldsymbol\lambda}(s;0) &=& 
a^{(0)}_{\boldsymbol\lambda} \Psi_{\boldsymbol \lambda}^{(0)}(s).
\end{eqnarray}
Substituting for $a^{(0)}_{\boldsymbol\lambda}$ and 
$\Psi_{\boldsymbol \lambda}^{(0)}(s)$, we can re-express it as
\begin{eqnarray}
 \label{gf-n2}
F_{\boldsymbol\lambda}(s;0) &=& 
\mbox{e}^{-\frac{k_+}{k_-}} \mbox{e}^{s\alpha_{\boldsymbol\lambda}}\nonumber\\
&=& \mbox{e}^{-\frac{k_+}{k_-}} \sum_{n=0}^\infty \frac{s^n}{n!} 
\alpha_{\boldsymbol\lambda}^n\nonumber\\
&=& \mbox{e}^{-\frac{k_+}{k_-}} \sum_{n=0}^\infty \frac{s^n}{n!}
\left(\frac{k_+^{(1)}}{k_-}\right)^n \nonumber\\
&\times&\sum_{m=0}^n \binom{n}{m}
\left(\frac{k_+^{(2)}}{k_+^{(1)}}\right)^m
\mbox{e}^{(n-m)\lambda_1} \mbox{e}^{m\lambda_2}.
\end{eqnarray}
Since $m,n$ are dummy variables, we can rewrite the last
line as
\begin{eqnarray}
 \label{gf-n3}
F_{\boldsymbol\lambda}(s;0) &=& 
\mbox{e}^{-\frac{k_+}{k_-}} \sum_{n=0}^\infty \sum_{N_2=0}^n
\frac{s^n}{n!}
\left(\frac{k_+^{(1)}}{k_-}\right)^n \nonumber\\
&\times&\binom{n}{N_2}
\left(\frac{k_+^{(2)}}{k_+^{(1)}}\right)^{N_2}
\mbox{e}^{(n-N_2)\lambda_1} \mbox{e}^{N_2\lambda_2}.
\end{eqnarray}
In order to put it in a convenient form which will allow an easy 
comparison with (\ref{eq-6}), we introduce a Kronecker delta
$\delta^K_{n-N_2,N_1}$. This allows us to rewrite Eq. (\ref{gf-n3})
as,
\begin{eqnarray}
\label{gf-n4}
F_{\boldsymbol\lambda}(s;0) &=& 
\mbox{e}^{-\frac{k_+}{k_-}} \sum_{n=0}^\infty \sum_{N_2=0}^n 
\sum_{N_1=-\infty}^{\infty}
\frac{s^n}{n!}
\left(\frac{k_+^{(1)}}{k_-}\right)^n \nonumber\\
&\times&\binom{n}{N_2}
\left(\frac{k_+^{(2)}}{k_+^{(1)}}\right)^{N_2}
\mbox{e}^{N_1\lambda_1} \mbox{e}^{N_2\lambda_2} \delta^K_{N_1,n-N_2}.
\end{eqnarray}
Finally, rearranging the Kronecker delta and using the fact that 
due to the binomial coefficients all terms for $N_2<0$ and 
$n<N_2$ vanish, we can recast this expression as
\begin{eqnarray}
\label{gf-n4}
F_{\boldsymbol\lambda}(s;0) &=& 
\mbox{e}^{-\frac{k_+}{k_-}} \sum_{n=0}^\infty \sum_{N_2=-\infty}^\infty 
\sum_{N_1=-\infty}^{\infty}
s^n \mbox{e}^{\lambda_1N_1} \mbox{e}^{\lambda_2N_2}
\nonumber\\
&\times& 
\frac{1}{N_1!N_2!} 
\left(\frac{k_+^{(1)}}{k_-}\right)^n 
\left(\frac{k_+^{(2)}}{k_+^{(1)}}\right)^{N_2}
\delta^K_{n,N_1+N_2}.
\end{eqnarray}
Comparing this with (\ref{eq-6}), we recover (\ref{distribution}).
Similar steps can be followed to obtain the time dependent 
joint distribution function given in Eq. (\ref{eq-17x2}).


\end{document}